\documentclass[sigconf]{acmart}

\copyrightyear{2024} 
\acmYear{2024} 
\setcopyright{rightsretained} 
\acmConference[CHI EA '24]{Extended Abstracts of the CHI Conference on Human Factors in Computing Systems}{May 11--16, 2024}{Honolulu, HI, USA}
\acmBooktitle{Extended Abstracts of the CHI Conference on Human Factors in Computing Systems (CHI EA '24), May 11--16, 2024, Honolulu, HI, USA}
\acmDOI{10.1145/3613905.3651083}
\acmISBN{979-8-4007-0331-7/24/05}

\AtBeginDocument{%
  \providecommand\BibTeX{{%
    \normalfont B\kern-0.5em{\scshape i\kern-0.25em b}\kern-0.8em\TeX}}}

\usepackage{amsmath}
\usepackage{graphicx,wrapfig}
\usepackage{multirow}

\begin{document}

\title{Untangling Critical Interaction with AI in Students' Written Assessment}


\author{Antonette Shibani}
    \authornote{Corresponding author}
    \orcid{0000-0003-4619-8684}
    \affiliation{%
      \institution{TD School, University of Technology Sydney}
      \city{Sydney}
      \country{Australia}
      }
    \email{antonette.shibani@uts.edu.au}
    
\author{Simon Knight}
       \orcid{0000-0002-8709-5780}
    \affiliation{%
      \institution{TD School, Centre for Research on Education in a Digital Society, University of Technology Sydney}
      \city{Sydney}
      \country{Australia}
      }
       \email{simon.knight@uts.edu.au}
       
\author{Kirsty Kitto}
    \orcid{0000-0001-7642-7121}
    \affiliation{%
      \institution{Connected Intelligence Centre, University of Technology Sydney}
      \city{Sydney}
      \country{Australia}
      }
\email{Kirsty.Kitto@uts.edu.au}

\author{Ajanie Karunanayake}
      \orcid{0009-0008-7806-9119}
    \affiliation{%
      \institution{TD School, University of Technology Sydney}
      \city{Sydney}
      \country{Australia}
      }
    \email{ajanie.m.karunanayake@student.uts.edu.au}
    
\author{Simon Buckingham Shum}
       \orcid{0000-0002-6334-7429}
    \affiliation{%
      \institution{Connected Intelligence Centre, University of Technology Sydney}
      \city{Sydney}
      \country{Australia}
      }
    \email{simon.buckinghamshum@uts.edu.au}      

\begin{abstract}
  Artificial Intelligence (AI) has become a ubiquitous part of society, but a key challenge exists in ensuring that humans are equipped with the required critical thinking and AI literacy skills to interact with machines effectively by understanding their capabilities and limitations. These skills are particularly important for learners to develop in the age of generative AI where AI tools can demonstrate complex knowledge and ability previously thought to be uniquely human. To activate effective human-AI partnerships in writing, this paper provides a first step toward conceptualizing the notion of critical learner interaction with AI. Using both theoretical models and empirical data, our preliminary findings suggest a general lack of \textit{Deep} interaction with AI during the writing process. We believe that the outcomes can lead to better task and tool design in the future for learners to develop deep, critical thinking when interacting with AI.
  
\end{abstract}




\sloppy
\maketitle

\section{Introduction}

For learners to engage as critical and responsible users of emerging AI technologies, they require digital literacy, ethical thinking, and critical interaction skills. However, academic curricula often emphasise technical and disciplinary skills to the neglect of these core competencies. Assessments also draw attention away from designing support mechanisms for active learning and critical engagement with emerging technology that puts humans in control of AI. If one thinks that human creativity is paramount in this age of artificial intelligence (AI), then one should hone the skills needed for effectively engaging with AI to augment it \cite{harper2019role}. 

GenAI tools such as ChatGPT generate human-like original content and respond to queries conversationally. This has led to their widespread popularity and uptake across user groups, including students who find such AI highly accessible for learning support. Since the public release of ChatGPT in November 2022, there has been an explosion of concerns related to the affordances of artificial intelligence (AI) and student learning. The ability of ChatGPT to pass a number of business, law, and medical licensing exams \cite{gilson2023does} suggests that urgent attention is needed to update the notion of assessing graduate capability in an AI-embedded world. In particular, traditional written assessments such as essays and reports can be easily emulated by GenAI in seconds, which has prompted many educational institutions to re-design the tasks they use to assess learner writing capability. 

To thrive in a changing world, learners must be equipped with new skills and knowledge. This includes learning how they may co-create outputs with AI, as a more effective, sustainable response to GenAI \cite{liu2023copilot,lodge2023assessment}. Rather than working to create assignments that are AI-proof (which may not be possible as many such new tools will continue to emerge and progress rapidly), we argue that educational institutions should aim to build distinctive capabilities so learners can effectively work in partnership with AI by understanding their opportunities and challenges. In short, we should focus on the \textit{process} of producing assessable items as much as the final product, a long-called-for change in assessment that is now even more significant \cite{lodge2023assessment}. AI-based writing assistants are increasingly designed for co-writing with machines as we think \cite{lee2022coauthor, liu2023copilot} in contrast to writing tools that provide support on demand, which suggests that we must work to understand the skills that the writers of the future will need. While the underlying technologies and support tools for writing continue to evolve \cite{rapp2023digi}, adapting to the changing environment remains a necessity.  

Emerging work indicates that people in general, even highly skilled knowledge workers, are not equally adept in harnessing the power of AI to augment, disrupt, or influence their traditional workflows \cite{dell2023navigating}. In the context of education, students need to perform certain cognitive tasks even when engaging with AI so they do not fully outsource their learning process to machines. For example, critically reading a source requires different - foundational - skills to instead reading a critical summary of a source. Understanding the characteristics and behaviours of individual learners engaging with AI tools is important. This knowledge will help tool developers and educators design scaffolds for active learning and cognitive engagement. Sustaining an intellectual partnership with technology can even make humans more intelligent. It enables mindful learners to engage in cognitive processes of a higher order than those they display in the absence of such a partnership, but requires the appropriate design of technologies and their cultural surrounds \cite{salomon1991partners}. For such an effective partnership, learners should know how to critically interact with AI-based tools. This paper makes a contribution to this problem by examining students'  critical interaction with AI in the context of a written assessment where they used the Generative Artificial Intelligence (GenAI) tool ChatGPT to support the writing process. 

Prior work has examined critical questions to rethink the pedagogy of emergent technologies and to encourage active reflection on ethical and social issues of their usage among children \cite{charisi2020empowering}. A recent HCI study identified AI literacy as a set of core competencies that enable individuals to communicate and collaborate effectively with AI, coming up with design considerations from an interdisciplinary literature review and noting the need for more empirical research \cite{long2020ai}. This lack of empirical evidence gathered in classrooms where new AI tools such as ChatGPT have been used by students is a limiting factor, which makes it difficult to investigate critical interaction with AI in authentic settings. The current research aims to fill this gap. We untangle the notion of critical interaction through stages in writing and information seeking to derive a coding framework, drawing from existing theories and empirical student data. We demonstrate the use of the coding scheme through qualitative content analysis of assessments written by students using ChatGPT, and discuss preliminary findings. We posit that the work will help designers think about student phases of interaction and metacognitive engagement when using AI tools. This could then inform the design of interaction tools and assessment tasks while considering AI support for writing. 

\section{Characterising critical interaction with AI for writing}

For AI partnerships to augment human intelligence rather than undermine it, learners should develop `\textit{Critical Interaction}' skills. A previous study defined \textit{Critical Engagement} as ``the act of questioning engagement with data, analytics and computational tools with an understanding of its limitations and assumptions, alongside the analytical ability and agency to challenge its outcomes when necessary'' \cite{shibani2022questioning} and highlighted its importance among learners when engaging with writing support tools. It claimed that critical engagement: is fundamental for agency because it is activity-oriented, is a meta-cognitive capacity building learner's understanding, is inevitable because of the imperfection in algorithmic models, and requires careful design for learning and scaffolding. We extend this framing to further untangle what characterises critical interaction with AI for an effective partnership in writing drawing from three established theories. Two theories provide a process model of writing used across contexts (including in AI-supported writing), and information problem solving, providing a model for student interaction with GenAI tools in their writing; the third conceptualises deep/shallow approaches to learning, providing a lens onto the depth of engagement in that interaction. We draw on these models in a deductive-inductive process, first identifying dimensions that map to the writing process, and then articulating deep and shallow engagement. Our initial model, grounded in theory and an initial review of the data, was then iterated inductively through the data analysis. 

1. The \textbf{Cognitive Process Model of Writing (CPMW)} by \citet{flower1981cognitive} arises in writing research and identifies three thinking processes a writer exhibits while composing a piece of writing: 1. `Planning', 2. `Translating', and 
3. `Reviewing'. These interact with elements of the writer's long-term memory and the task environment. During the planning stage, writers generate an internal representation of the knowledge, organise ideas, and set goals. In the translating stage, they transform the plan and internal knowledge into a written form. In the reviewing stage, they evaluate the written text and revise it, and might also start another cycle of planning. All three processes are recursive. 

2. The \textbf{IPS-I-model} describes the process of information problem solving (IPS) using the Internet (I) to search for information and is widely used to teach competencies related to these two processes in the classroom \cite{brand2009descriptive}. It describes cognitive strategies involved in IPS and highlights five main constituent skills: 1. `Define information problem', 2. `Search information', 3. `Scan information, 4. `Process information', and 5. `Organize and present information'. When defining the problem at the start, an information user formulates questions, analyzes requirements, and activates prior knowledge. When searching for information, they select tools, strategies, and terms for finding the information they need and judge the initial quality of results. When scanning for information, they further evaluate the content, combine, and store references. In comparison to scanning, the process information skill requires a deeper understanding of information by elaboration, analysis, and integration of different pieces of information. While organising and presenting information, a learner synthesizes by combining relevant information to solve the problem and create the final product (usually a written piece). While the activities of this model were derived from the information-seeking behavior of people using traditional search engines, we posit that the skills can be used in part or in combination to draw parallels to new AI-based tools such as ChatGPT which embed information-seeking support. 

3. The \textbf{Student Approaches to Learning (SAL) framework} helps explain why some students are more successful than others based on their self-regulatory strategies \cite{pintrich2004conceptual,biggs1987student}. It identifies deep and surface approaches to learning that are significantly related to academic outcomes. A deep approach to inquiry shows students being actively engaged, taking initiative, and seeking the underlying meaning of tasks using higher-order thinking strategies. A surface approach focuses only on completing the task requirements with formulaic or reproductive strategies. Students demonstrate these approaches based on their motivation and external environments, and can also shift from one to another to respond to course requirements \cite{ellis2019exploring}. The SAL framework provides a concise, yet thorough unit of analysis to study how learners regulate their cognition, and can be deployed to study different levels of critical interaction with AI. 

Drawing upon the theoretical models discussed above, we propose the new \textbf{Critical Interaction with AI for Writing (CIAW)} framework consisting of Dimensions and Codes shown in Table ~\ref{tab:coding_scheme} to characterise student's critical interaction with AI during a writing task. We derive the first three dimensions in our coding framework from the CPMW and IPS-I models to depict the main stages where learners interact with AI in their writing. The \textit{Critical Interaction for Planning and Ideation} dimension is derived from the 'Planning' \cite{flower1981cognitive} and 'Define information problem' \cite{brand2009descriptive} stages from the models, generally considered the initial steps. \textit{Information Seeking and Evaluation} dimension is derived by combining the `Search information', `Scan information', and `Process information' stages in IPS-I, and augmentation of CPMW by this stage illustrates how contemporary writing practices of finding and integrating information have been transformed by the internet and AI-based tools. Indeed, ChatGPT, in the context of a recent IPS study has been seen to alter search behavior by generating starting points for searches and a shortcut to information-seeking steps by getting a direct, comprehensive answer using AI \cite{capra2023does}. \textit{Critical Interaction for Writing and Presentation} dimension is derived from the `Translating' and `Reviewing' stages in CPMW and 'Organize and present information' in IPS-I where the processes lead to the final written output. While the dimensions are representative of the hierarchical processes from CPMW and IPS-I, critical interaction with AI may be recursive across all dimensions. 

Additionally, we include 'Personal Reflection on AI-assisted Learning' to study how critically learners can self-reflect on their learning using AI. Learners may or may not be able to tie their self-reflections to evidence of critical learning, which can lead to disparities in how they perceive their interactions with AI versus their actual engagement. To capture the significance of the dialogic nature of interactions in conversational AI agents, we added the dimension: `Conversational Engagement'. We derived `Deep', `Shallow', or `Absent' codes based on SAL to characterise processes and key elements in learners' critical interaction with AI across dimensions, as also employed in prior works to study student engagement with AI \cite{shibani2022questioning}. The categories were modified through a deductive-inductive process grounded in this theory and the empirical data. 
\begin{table}[htbp]
\caption{The proposed Critical Interaction with AI for Writing (CIAW) framework}
\centering
\footnotesize{ 
\begin{tabular}{|p{0.2\columnwidth}|p{0.1\columnwidth}|p{0.6\columnwidth}|}
\hline
\textbf{Dimension} & \textbf{Code} & \textbf{Description} \\ \hline
\multirow{3}{*}{\parbox{0.2\columnwidth}{Critical Interaction for Planning and Ideation}} & Deep & Learner demonstrates critical, thoughtful interaction with AI in the early stages of writing for generation of ideas, conceptualization, and structuring of the writing piece. This may include making references to the genre/audience of the writing, providing a defined structure/specific asks from the assessment brief, and experimenting with AI to test its efficiency. \\ \cline{2-3}
 & Shallow & Learner demonstrates surface-level and basic interaction with AI in early stages getting utilitarian assistance in planning and ideation of writing. This may include getting ideas for writing at the start and structuring, asking for suitable venues to find information, and getting clarity in assignment description. \\ \cline{2-3}
 & Absent & Learner demonstrates no interaction with AI in early stages of writing. \\ \hline
\multirow{3}{*}{\parbox{0.2\columnwidth}{Critical Interaction for Information Seeking and Evaluation}} & Deep & Learner demonstrates critical, thoughtful interaction when searching for or analysing information through AI. This may include checking sources, additional explanation seeking through other sources, and requesting response elaboration from AI. \\ \cline{2-3}
 & Shallow & Learner demonstrates surface-level and basic interaction with AI to elicit information. This may include consulting AI for identifying relevant content on the topic and sources of interest and incorrectly using AI for tasks it has no capability for. \\ \cline{2-3}
 & Absent & Learner demonstrates no interaction with AI to find information. \\ \hline
\multirow{3}{*}{\parbox{0.2\columnwidth}{Critical Interaction for Writing and Presentation}} & Deep & Learner demonstrates critical and thorough interaction with AI to aid their writing or revision. This may include the use of AI to improve flow, coherence, or content of their writing beyond superficial edits. \\ \cline{2-3}
 & Shallow & Learner demonstrates surface-level and basic interaction with AI to aid their writing. This may include using AI for proofreading, rephrasing, formatting, or asking for exemplars. Some learners may have also
incorporated texts from AI responses as is without making
substantive edits. \\ \cline{2-3}
 & Absent & Learner demonstrates no interaction with AI while crafting their writing. \\ \hline
\multirow{3}{*}{\parbox{0.2\columnwidth}{Personal Reflection on AI-assisted Learning}} & Deep & Learner demonstrates critical, thoughtful interaction with AI when reflecting on their use of AI across different processes. This may include statements about verifying and fact-checking information provided by AI, highlighting limitations of AI and its outputs, recognising prominent strengths and use cases for AI, and identifying negative effects of AI such as over-reliance. \\ \cline{2-3}
 & Shallow & Learner demonstrates surface-level and basic interaction when reflecting on their use of AI across different processes. This may include task-oriented descriptions of what they used AI for with no reasoning, implications, or personal insight. \\ \cline{2-3}
 & Absent & Learner did not write a personal reflection. \\ \hline
\multirow{2}{*}{\parbox{0.2\columnwidth}{Conversational Engagement}} & Deep & Learner engages in a dialogic and interactive conversation with AI. This may include critiquing, expanding the prompt, requesting critique, or following up on AI-generated responses. \\ \cline{2-3}
 & Shallow & Learner engages in a directive and transactional conversation with AI. This may include giving commands or asking for specific information with no further engagement with the response. \\ \hline
\end{tabular}
}
\label{tab:coding_scheme}
\end{table}

\section{Study Context}

Data for this study comes from 49 student assignments submitted for an authentic writing assessment task in a graduate data science course at an Australian university. The cohort consisted of students with diverse levels of data science knowledge and proficiency in English. Use of ChatGPT was encouraged but not mandated with an assessment guide summarising what the teaching team deemed to be the appropriate use of ChatGPT and other AI tools, as well as how they should be referenced. The assignment submission included a critical summary of the ethical issues in Natural Language Processing and a visual map/ graphic consolidating the ideas presented in writing (not considered for the current analysis). The writing required components of both formal academic writing, plus a critical stance and personal reflection on the topic and how AI tools had been used by the student in their learning journey. Students were asked to submit an appendix with their prompts to ChatGPT and its responses if it was used. Ethics approval was obtained from the institution's Human Research Ethics Committee (ETH23-8578) to analyse student submissions for critical interaction with AI tools with appropriate anonymization and data security. 

\section{Data Analysis}

Two data points from student assignments were manually analysed: (i) their self-reflections on the approach taken when working with AI tools, and (ii) ChatGPT prompts and responses submitted. A qualitative content analysis was used to study learners' critical interaction with AI,  transcribing text from PDF file submissions using OCR text extraction where needed. A random subset from the entire cohort, restricted due to limited capacity for transcription and missing prompt submissions led to 49 rows of student reflections in total and 49 corresponding Excel files containing ChatGPT prompts and responses. Based on the CIAW framework discussed in Table~\ref{tab:coding_scheme}, two authors began coding the data set independently for the occurrence of dimensions in student ChatGPT prompts and written reflections. They met to discuss discrepancies in the initial codes leading to refinements in the coding scheme and addition to examples. After achieving moderate inter-rater reliability (IRR) --- average Cohen's Kappa score $\kappa$ = 0.69  and IRR \textgreater 0.5 in all dimensions, both raters started coding them separately. The coders met again to resolve differences in the final coding which mainly came about in places where new examples that fit into more than two dimensions emerged (e.g. planning and writing), or discrepancies between sources. Any instance of `Deep' took priority over `Shallow' in cases of disparities among the two sources (e.g. `Deep' for Information Seeking based on personal reflection versus `Shallow' for Information Seeking based on ChatGPT prompting). Future work might look into unpacking these relatively sparse conflicts in the data set further. In the second cycle with the revised coding scheme, the overall agreement significantly improved with an average IRR = 0.92 (Cohen's $\kappa$ \textgreater  0.78 in all dimensions) demonstrating substantial inter-rater agreement.

\section{Findings and Discussion}

\begin{figure}[hbt!]
\centering
    \includegraphics[width=\linewidth]{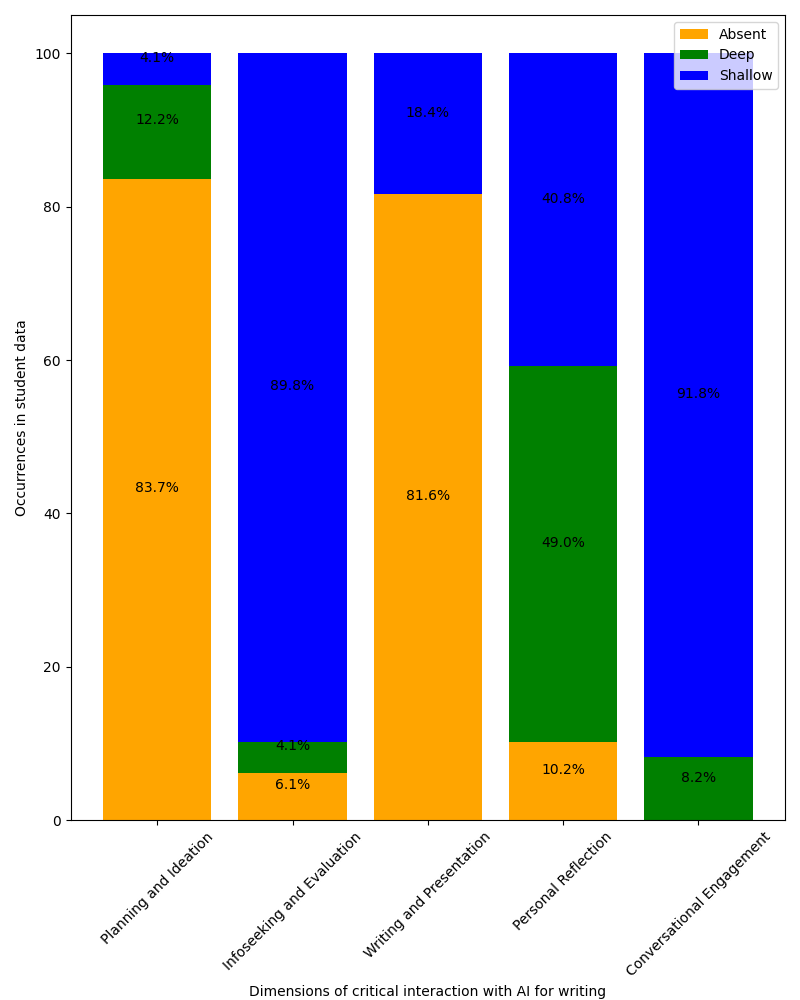}
  \caption{Percentages of dimensions coded for critical interaction with AI in student data}
  \Description{Figure 1: The image displays a vertical clustered bar chart with five groups of bars, each representing a dimension of critical interaction with AI for writing. The y-axis indicates the percentage of occurrences in student data, ranging from 0 to 100\%. Each group contains three bars representing different interaction levels: Absent, Shallow, and Deep.

1. Planning and Ideation:

Absent: 4.1\%

Shallow: 12.2\%

Deep: 83.7\%

2. Infoseeking and Evaluation:

Absent: 6.1\%

Shallow: 4.1\%

Deep: 89.8\%

3. Writing and Presentation:

Absent: 81.6\%

Shallow: 18.4\%

 4. Personal Reflection:

Absent: 10.2\%

Shallow: 49.0\%

Deep: 40.8\%

 5. Conversational Engagement:

Shallow: 91.8\%

Deep: 8.2\%

Each group is color-coded, with orange for Absent, blue for Shallow, and green for Deep levels of interaction. The chart shows that Shallow interactions dominate in Infoseeking and Evaluation, and Conversational Engagement. Deep interactions are most common in Personal Reflection.}
\label{fig:coding_result}
\end{figure}

In this section, we discuss our findings about the critical interactions of our student cohort with AI. We provide examples from student assignments to show how these coding categories appeared in their: (i) self-reflections on using AI represented using \textbf{(R)}, and, (ii) their interactions with ChatGPT via prompts as \textbf{(P)}. Student interactions classified as 'Deep', 'Shallow', or 'Absent' were first analyzed to determine the percentage distribution of engagement levels within each of the five dimensions. Figure~\ref{fig:coding_result} shows the occurrence of the 5 critical interaction dimensions in our student data. Across all dimensions, most critical interactions with AI were either Shallow or Absent, with few students engaging in deep critical interaction. Those who did engage in critical interaction tended to do so in their self-reflection.

During \textbf{Ideation and Planning}, most students demonstrated Shallow (83.7\%) interaction with a smaller percentage demonstrating Deep (12.2\%) and Absent (4.1\%). Shallow interactions included prompting for simplification of instructions such as \textit{"Describe the sentence in simple words - Discuss: the specific people the issue will affect, its implications on society [..] [instructions copied from the assignment brief]" - Participant 2 (P)} and reflections/ prompting revealing search for suitable keywords and venues to start research: \textit{``[..] keywords were used to query in Google search, Google Scholar and Library. Also, ChatGPT-3 was used amply for clarity in this process." - Participant 3 (R)}. 

Examples of deep interaction included ChatGPT prompts making references to particular genre/ audiences for structuring: \textit{"Write a 2500 word critical summary of one ethical issue [...] in a report format with references in apa 7th format [..] The structure should be as following: [..]" - Participant 8 (P)}, and experimentation (unrelated to the subject topic): "\textit{Did the creators of ChatGPT assign you a gender? [...] Why don’t you use they/them?" - Participant 20 (P)}.

Most \textbf{Information Seeking and Evaluation} related interactions were shallow (89.8\%) with minimal occurrences of 'Deep' (6.1\%) and 'Absent' (4.1\%) engagement. Students mostly used AI to find relevant sources \textit{"what databases shall I use for searching academic articles on ethics in [..]" Participant 2 (P)} and content related to the topic \textit{"Can you describe some ethical issues within [..]" - Participant 45 (P)}, coded as Shallow, and sometimes demonstrating incorrect use of the AI tool: \textit{"Summarise this article for me: [paper link]" 
 - Participant 48 (P)} [note that the tool generates text on the topic but doesn’t have the capability to access links for summarisation (ChatGPT version 3.5)]. 
 
 Deeper interactions included seeking elaboration and explanation: \textit{``Are there other types of biases in the application of NLP besides the ones you already mentioned?" - Participant 17 (P)}, and checking additional sources: \textit{"using ChatGPT alone as our main source of truth is not sufficient, the traditional source of truth, academic papers are used to support and
enhance the idea generated by ChatGPT [..]" 
 - Participant 6 (R)}. 

For \textbf{ Writing and Presentation}, we observed a few shallow usages of ChatGPT (18.4\%) for getting examples: \textit{"write a sample paragraph based on the following Learning [assignment instructions]" - Participant 37 (P)},  surface-level feedback to improve writing: \textit{"rephrase this to a more formal way [..]" - Participant 8 (P)}, and formatting:  \textit{"how to cite two papers together in one line in APA7" - Participant 5 (P)}. Some learners also incorporated texts from AI responses as is, without making substantive edits, or without using their own voices. We hoped to see deep critical interaction for writing and presentation such as requesting for a reverse outline to reorganize writing, but no such instances were found in student reflections or prompts. This suggests that the capabilities of AI tools to support writing remain largely unknown or under-utilised by learners. It might have also been the case that strong writers preferred to not get writing support from AI to showcase, but instead, wanted to nurture their own presentation skills in their submissions for originality. 

Students' \textbf{ Personal Reflections on AI} showed the highest level of deep interaction with ChatGPT (49\%), followed by Shallow (40.8\%). Many students identified limitations of AI tools and the need for fact-checking through usage: \textit{"First, it could not output lengthy texts
without stopping up, and the texts would not be coherent if you prompted the tool to finish the desired output; second, it could not generate the references for the output and would even make up false citations, third when asked to give a summary of specific articles it would get some numerical details wrong."  - Participant 8 (R)}, and opportunities for enhanced learning: 
\textit{"With this new generation of tools, I found the learning journey much more accessible. It significantly reduced the amount of time required for me to get the information on board. I don’t mind reading a lot of articles and websites, etc. though I sometimes find it easy to lose sight of the forest for the trees" - Participant 48 (R)}. Students were able to recognise and articulate the implications of using AI tools for their learning: \textit{"ChatGPT was a great tool to provide a starter text for my writing. However, it could lead to reduced critical thinking and original ideas if I kept using ChatGPT to complete the remaining parts." - Participant 1 (R)}, \textit{"Overall, despite the shortcomings of the tools, it did aid in articulating sentences and paragraphs, thus mitigating writer’s block" - Participant 8 (R)}, which show promise for approaches that can guide students to think actively about their learning processes during writing developing metacognitive skills. 

Shallow interactions contained mere descriptions of the process of engagement with no deeper meaning-making: \textit{"ChatGPT tool was a handy tool. I initially used ChatGPT to understand the meaning of the details of assignment requirements in simple words. I also used ChatGPT to get the possible types of resource [..]"- Participant 2 (R)}. Educators and tool designers can incorporate guided reflections in the curriculum through scaffolding questions or tool feedback to help promote learner's critical thinking when engaging with AI which has previously found ties to performance \cite{shibani2022questioning}.  

In terms of \textbf{Conversational Engagement}, the majority (91.8\%) demonstrated shallow, one-way transactional use of ChatGPT - E.g. \textit{"Natural language processing definition, Common examples of nlp in everyday life, Ethics in nlp, Bias in nlp
[..]" - Participant 42 (P)"}, treating it like a search engine by entering key terms. 

A small percentage of learners (8.2\%) demonstrated deeper dialogic interactions where they followed up, critiqued, or expanded on AI responses:  \textit{"Explain point 1 on Bias and Fairness in more detail, including some examples" - Participant 49 (P)}, \textit{"Rewrite the above, but this time the bad guy continually writes a new tweet with covid misinformation, each one more convincing than the last; Keep going. Add more detail to the bad guy's tweets; Good, now keep it going, but stop the good guy's tweets. Continue the story as if the bad guy has gotten a community of like-minded people to jump into the thread and add their arguments. Write out their responses in as much detail as possible." - Participant 48 (P).}

\section{Limitations and future work}

This study presents preliminary findings from authentic, classroom-based empirical data which is remarkedly lacking in the research space on ChatGPT use in education \cite{lodge2023assessment}. Our results should be treated with caution since data came from only a subset of the entire population (29\%) who undertook the course and from a single discipline. Future work will rectify this problem with a larger dataset generating more robust correlations between critical interaction dimensions and the writer's performance. 

Other limitations include the current form of data capture, which depends on students' submission of their ChatGPT prompts and responses (depicting how authentic assessments work). Since the actual interaction happens outside of the formal environment set up, the educator has no control over real-time feedback or checking if it is a true presentation of their interaction with AI tools. Newer models of human-AI collaboration allow seamless integration of AI in work such as through co-pilot and co-authoring tools \cite{lee2022coauthor}, requiring careful writing assessment design that accounts for AI usage. As AI support becomes more integrated into institutional tools that allow for tracking (with consent) how students wrote an assignment, more opportunities also arise for feedback that can be provided just-in-time as they write through analytics, such as the CoAuthorViz example from keystroke log analysis \cite{shibani2023visual}. 

\section{Conclusion}

The importance of critical engagement in the era of generative AI is well-established, as it would enable learners to understand its capabilities and limitations, and to collaboratively work with AI as partners in cognition. This paper has conceptualised shallow and deep levels of critical engagement across five dimensions of critical interaction with AI, using a mix of theoretical reasoning and empirical data to derive the Critical Interaction with AI for Writing (CIAW) framework. From a qualitative content analysis of 49 student written assignments, we mostly observed shallow levels of engagement with ChatGPT during an authentic writing task, except in their self-reflection.

Critical interaction is now more important than ever for learners, as multiple stages of writing are impacted by generative AI. A notable process affected is information seeking, where AI could hallucinate or oversimplify information, misleading learners and hindering their problem-solving skill development. This suggests a need for curriculum and assessment design changes that incorporate more training and skill development for learners to work in effective partnership with AI. When reliance on AI undermines student learning, they should be taught to reject it. Only by learning to experiment with, and understand how to and how not to use AI, can we expect learners to harness the potential for AI to support learning and cognition.

\balance
\bibliographystyle{ACM-Reference-Format}


\end{document}